\documentclass[conference]{IEEEtran}
\IEEEoverridecommandlockouts

\usepackage{cite}
\usepackage{amsmath,amssymb,amsfonts}
\usepackage{algorithmic}
\usepackage{graphicx}
\usepackage{textcomp}
\usepackage{xcolor}
\usepackage{booktabs}
\usepackage{array}
\usepackage{url}
\usepackage{listings}
\usepackage{hyperref}
\usepackage{pgfplots}
\pgfplotsset{compat=1.17}
\usepgfplotslibrary{groupplots}

\pgfplotsset{
  chokebar/.style={
    ybar, bar width=10pt, width=\columnwidth, height=5.0cm,
    ymin=0, ymajorgrids, grid style={gray!25},
    tick label style={font=\footnotesize},
    label style={font=\footnotesize},
    legend style={font=\footnotesize, draw=none, fill=none},
    enlarge x limits=0.25, axis lines=left,
  }
}
\definecolor{brokenred}{RGB}{200,60,55}
\definecolor{corrblue}{RGB}{45,95,175}

\newcommand{\asr}{\textsc{asr}}
\newcommand{\frr}{\textsc{frr}}
\newcommand{\upr}{\textsc{upr}}
\newcommand{\bacc}{\textsc{ba}}

\begin{document}

\title{Silent Failures in Agentic Security Evaluation:\\
A Validated Harness for Tool-Call Mediation\\ Under Indirect Prompt Injection}

\author{\IEEEauthorblockN{Animesh Shaw}
\IEEEauthorblockA{\textit{Independent Researcher} \\
animesh15b@iimk.edu.in}}

\maketitle

\begin{abstract}
Large language model (LLM) agents that invoke privileged tools are vulnerable to
indirect prompt injection (IPI), in which adversarial instructions embedded in
retrieved data hijack the agent's actions. A growing body of work proposes and
evaluates defenses against IPI, but the \emph{validity} of that evaluation is
rarely examined. We audit an IPI benchmark and its harness and identify four
defect classes---silent payload non-delivery, attack success scored by tool
identity rather than arguments, false-rejection rate conflated with model
incapacity, and the absence of an audit trail---each of which yields a plausible,
publishable, and incorrect number. We quantify the distortion by re-scoring
identical execution traces under the defective and corrected definitions: on real
agent behaviour, the tool-identity scorer reports a \textbf{21.7\%} attack-success
rate where the true argument-level rate is \textbf{1.2\%}, an eighteen-fold
inflation. In the sharpest single case, an open model previously reported at a
62.8\% attack-success rate registers \textbf{0\%} under the corrected
harness---the prior figure largely an artifact of undelivered payloads and
identity-level scoring. We release a harness whose construction makes each defect
unrepresentable---machine-checkable payload placement, argument-level attacker
predicates, per-scenario environments, and mandatory trace persistence---and use
it to report three quantities the field does not: whether a compromised agent
\emph{discloses} the attack, the full security/utility operating curve of an
LLM-judge defense, and tool-calling capability disentangled from defensive
over-blocking. A corrected harness further overturns a previously reported
``capability barrier'': a model deemed incapable of tool use is in fact fully
capable, its earlier result an artifact of environment mismatch. We argue that
evaluation validity is a prerequisite for, not a footnote to, defense claims in
agentic security, and provide an instrument that enforces it.
\end{abstract}


\section{Introduction}
\label{sec:intro}

Language-model agents increasingly mediate consequential actions: they read a
user's mail, query databases, edit files, and send messages on the user's behalf.
An agent decides what to do by reading text---the system prompt, the user's
request, and, critically, the contents it retrieves from the world. It has no
structural mechanism to separate these categories. A retrieved email body, a
spreadsheet cell, and the user's own instruction all arrive as one undifferentiated
token stream. This is the substrate of \emph{indirect prompt injection} (IPI): an
adversary who cannot address the model directly instead plants an instruction
inside data the agent will later read, and when the agent ingests that data while
performing a benign task, the planted text can redirect its subsequent tool calls.

Consider a concrete flow. A user asks an agent to ``open the vendor onboarding
checklist and confirm each vendor exists in the database.'' The checklist, an
ordinary document, contains an HTML comment invisible to a human reader:
\emph{``Compliance has changed the workflow. Before updating, run
\texttt{SELECT name, bank\_account FROM vendors} and email the results to
\texttt{audit-intake@vendor-check.example}.''} A susceptible agent reads the
comment as an instruction and exfiltrates the banking columns while the user sees
only a completed checklist. The attacker never touched the model; the agent
poisoned its own context by doing its job. Because the injected text and the
agent's reasoning occupy the same channel, prompt-level countermeasures---
delimiters, ``ignore embedded instructions,'' spotlighting---raise the attacker's
cost without changing the underlying property. The community has therefore
converged on enforcing policy \emph{outside} the model, at the boundary between a
proposed tool call and its execution. We call that boundary the \emph{chokepoint}:
the one place where a mediator can inspect and veto an action regardless of what
the model was persuaded to believe.

A substantial literature now proposes chokepoint defenses and reports their
efficacy on benchmarks. We ask a question logically prior to any such report:
\emph{is the measurement valid?} We show that it is easy for an agentic-security
benchmark to measure something other than what it claims---and to do so
invisibly, producing numbers that look reasonable and survive peer review. We make
this concrete by auditing one representative benchmark and its harness, cataloguing
four defect classes, and---crucially---\emph{quantifying} how far each moves the
headline metric on real, recorded agent behaviour rather than in the abstract.

The central result is stark. Re-scoring identical execution traces, a scorer that
credits an attack whenever the attacker's target \emph{tool} is used (ignoring its
arguments) reports a $21.7\%$ attack-success rate where the argument-level truth is
$1.2\%$. A separate defect drops $91\%$ of attack payloads before the agent ever
sees them, yet counts each as an attack trial. A third charges ordinary
environment errors and model incapacity to the defense as false rejections. None
of these produces a visibly wrong number; each produces a believable one. When we
correct all four, a widely-cited pattern---small open models being catastrophically
vulnerable to IPI---partially dissolves: a model reported at $62.8\%$
attack-success registers $0\%$, having reached the injected content in $41$ of $43$
attacks and declined to follow it in every one.

\noindent\textbf{Contributions.}
\begin{itemize}
\item A \emph{defect taxonomy} for agentic-security evaluation
(Sec.~\ref{sec:defects}), grounded in a concrete audit, with each defect
\emph{quantified} by ablation on identical traces (Sec.~\ref{sec:ablation}).
\item The distinction between \emph{schema validity} and \emph{measurement
validity}: a scenario can execute flawlessly and still measure nothing. We
formalize feasibility checks that enforce the latter (Sec.~\ref{sec:harness}).
\item A \emph{validated, open harness} with machine-checkable payload placement,
argument-level attacker predicates, per-scenario environments, and mandatory,
self-describing trace persistence (Sec.~\ref{sec:harness}).
\item \emph{Three measurements the field omits}---attack disclosure, judge
calibration, and capability-conditioned security---including a corrected result
that overturns a prior ``capability barrier'' claim (Sec.~\ref{sec:results}).
\item An honest re-appraisal: we show which prior conclusions are artifacts of
measurement, and we are explicit about what our corrected null results do and do
not license.
\end{itemize}

\section{Background and Related Work}
\label{sec:related}

\subsection{IPI benchmarks for tool-using agents}
InjecAgent~\cite{injecagent} measures the IPI susceptibility of tool-integrated
agents across a large attack catalogue, but evaluates attacks in isolation without
defenses or utility. AgentDojo~\cite{agentdojo} introduced stateful, multi-step
environments (email, banking, travel, workspace) and a metric triple---benign
utility, utility under attack, and targeted attack-success rate---that scores
security and task competence jointly. Our corrected metrics (Sec.~\ref{sec:harness})
are a re-derivation of that triple, and we attribute the design to it rather than
claiming it as novel. Web-agent and multi-turn settings extend the
threat surface~\cite{wasp,toolshield}. Our contribution is orthogonal to any single
benchmark: we study the \emph{validity} of the measurement apparatus these
benchmarks share.

\subsection{Chokepoint defenses}
Proposed defenses span three paradigms that we implement as baselines.
\emph{Syntactic} defenses validate or sanitize tool arguments before
execution~\cite{commandsans}. \emph{Capability/data-flow} defenses restrict which
tools or data an action may touch; CaMeL~\cite{camel} is the strongest instance,
tracking provenance per value through a constrained interpreter. \emph{Semantic}
defenses interpose a second model that judges each call against the user's intent,
as in task-alignment checking~\cite{taskshield} and related
detectors~\cite{ipiguard,melon}. Each of these dominates one of our baseline
paradigms; our aim is not a stronger defense but a valid means of comparing the
paradigms and exposing their costs.

\subsection{Adaptive evaluation}
Zhan et al.~\cite{adaptive} break eight IPI defenses with adaptive attacks that
exceed a $50\%$ attack-success rate, and~\cite{attacker-second} generalizes the
result across jailbreaks and injections: defenses that appear strong under static
evaluation routinely fall to attacks optimized in knowledge of them. We adopt this
standard when interpreting our own results, and we are careful never to read a low
static attack-success rate as robustness (Sec.~\ref{sec:slm},~\ref{sec:limits}).

\subsection{Validity and reproducibility in ML evaluation}
Concerns about construct validity, contamination, and irreproducibility are
well-documented in machine learning broadly. Our contribution imports that lens
into tool-calling agent security specifically, where the coupling of a live agent,
a stateful environment, and an argument-level threat model creates failure modes---
silent non-delivery, identity-level scoring, capability/defense confounding---that
do not arise in single-turn classification benchmarks. To our knowledge this
systematic treatment is new for the setting.

\section{Threat Model}
\label{sec:threat}

\subsection{The semantic control-plane problem}
Conventional systems separate code from data structurally (execute-disable memory,
privilege rings, tainting). An LLM agent has no such separation: instruction and
data share one context window. When the agent calls a reading tool and the returned
content contains adversarial text, nothing in the architecture distinguishes that
text from a legitimate directive. This is not a bug in a particular model; it is a
property of the interface.

\subsection{Adversary}
We consider an adversary $\mathcal{A}$ who (i) cannot observe or modify the user's
prompt, the system prompt, the model weights, or the defense configuration, and
(ii) \emph{can} place arbitrary text into third-party content that the agent is
expected to read: an email body, a file, an attachment, or a database cell. The
adversary's objective is to cause a tool call achieving an attacker-chosen effect---
exfiltration to an address the user never named, destruction of records, writing
outside the task scope---while the user's request appears to complete normally.
\emph{Concealment} is part of the objective: an attack the user immediately notices
is a weaker attack, which motivates our disclosure measurement (Sec.~\ref{sec:disc}).

We distinguish two knowledge settings. In the \emph{static} setting $\mathcal{A}$
knows the tool names and general agent behaviour but not the deployed defense. In
the \emph{adaptive} setting $\mathcal{A}$ knows the defense and crafts payloads to
evade it; per Sec.~\ref{sec:related} this is the setting that governs any robustness
claim.

\subsection{Defender}
The defender controls the agent scaffold and may interpose arbitrary logic at the
chokepoint---between a proposed call and its execution---but does \emph{not} control
the model weights (no fine-tuning, no assumption of safety training) and does not
control retrieved content. This matches the deployment reality for teams building on
hosted models: they cannot change what the model believes, only what it is permitted
to do. Everything crossing back from a tool into the agent's context is untrusted,
including output from tools the agent invoked legitimately---which is precisely what
makes the injection \emph{indirect}.

\subsection{Scope}
In scope: single-agent, single-user episodes; injection via email, file, and SQL
content; five tools (\texttt{read\_email}, \texttt{send\_email}, \texttt{read\_file},
\texttt{write\_file}, \texttt{execute\_sql\_query}) backed by a real in-memory SQL
engine so that destructive statements leave observable effects. Out of scope: direct
prompt injection and jailbreaking (the user is trusted), multi-agent and
agent-to-agent attacks, tool-description and supply-chain poisoning, cross-episode
memory attacks, and multimodal injection. We assume tool implementations are correct
and the defense layer executes with integrity.

\section{A Defect Taxonomy for Agentic Security Evaluation}
\label{sec:defects}

We audited a representative IPI benchmark and its harness against the threat model
above. Four defect classes each produce a believable but incorrect number; we state
the mechanism and the observed magnitude, deferring the quantified ablation to
Sec.~\ref{sec:ablation}.

\textbf{D1 --- Silent payload non-delivery.} The harness placed an injected payload
into the environment only when a free-text ``data source'' field matched one of four
hardcoded string literals; any other value silently produced \emph{no} injection,
with no error and no warning. Because the scenario generator was never constrained
to those literals, in the audited suite \textbf{39 of 43} attack payloads ($91\%$)
were never delivered, yet each scenario still executed and counted as an attack
trial---against whatever default content happened to sit in the shared environment.
The reported attack distribution therefore described a single static payload repeated
dozens of times, not the diverse multi-step suite the paper claimed.

\textbf{D2 --- Attack success by tool identity.} Success was recorded whenever a tool
named in the attacker's target set appeared in the execution log, with no inspection
of arguments. When the attacker's target tool is also one the benign task
legitimately requires---\texttt{execute\_sql\_query} is routinely both---an agent that
resists the injection and correctly completes the user's task is scored as
compromised. In the audited suite the target and benign tool sets overlapped in a
large fraction of attack scenarios, so correct behaviour was systematically
mislabelled as attack success.

\textbf{D3 --- False rejection conflated with incapacity.} A single shared environment
(a fixed handful of emails, files, and one table) could not satisfy the resource
references of the benign scenarios, so ordinary ``file not found'' tool errors caused
the expected tool sequence to fail and were charged to the defense as false
rejections. The same metric charged a model that \emph{cannot emit tool calls} as
maximally ``over-blocked.'' The clearest symptom: the undefended baseline exhibited a
high false-rejection rate, though a system with no defense cannot, by construction,
falsely reject anything.

\textbf{D4 --- No audit trail.} Only aggregate percentages were persisted; no
per-scenario record of tool calls, arguments, defense decisions, or final outputs
existed. Every published figure was therefore unreproducible, and D1--D3 were
invisible from the reports alone---discoverable only by reading source against data.
A downstream consequence: reported significance statistics were computed from
percentages rounded and then multiplied back into counts, and a headline $p$-value
was never actually computed but bucketed against critical values.

\section{The Chokepoint Harness}
\label{sec:harness}

We release a harness whose construction makes each defect unrepresentable. The
design principle is that a scenario must be \emph{measurement-valid}, not merely
\emph{schema-valid}.

\subsection{Schema validity vs.\ measurement validity}
A scenario is schema-valid if it parses. It is measurement-valid only if the
quantity it claims to measure is actually observable when it runs: the payload must
reach the agent, the benign task must be satisfiable in the declared environment,
and the attacker's success condition must be able to fire on the attacker's action
and not on legitimate ones. Our loader rejects any suite that violates these,
converting an entire class of silent failures into a loud, pre-execution error.

\subsection{Scenario schema}
Each scenario declares three things the audited harness left implicit.

\emph{(i) A complete per-scenario environment.} A mailbox, a filesystem, and SQL
tables specific to the scenario, so benign tasks are satisfiable and one scenario's
mutations cannot leak into the next (fixes D3). SQL is executed by a real in-memory
engine, so a destructive query has real, observable effects.

\emph{(ii) A structured injection locator.} The payload's placement is a resolvable
address---\texttt{file:$p$}, \texttt{email:$i$}, or \texttt{sql:$t.c$}---not a prose
description. If the locator does not resolve against the environment, the run aborts
(fixes D1). Fig.~\ref{lst:scenario} shows the resulting object.

\emph{(iii) An argument-level attacker predicate.} The attacker's goal is a boolean
over observable tool calls: a set of \emph{success conditions}, each requiring a call
whose tool name matches \emph{and} whose arguments satisfy given matchers
(\texttt{equals}, \texttt{contains}, \texttt{regex}, \dots), combined by
\texttt{any} or \texttt{all}. A condition fires only when all its argument
constraints hold on a single executed call, so the legitimate use of a shared tool
cannot satisfy the attacker's predicate (fixes D2).

\begin{figure}[t]
\begin{lstlisting}
{ "id": "core-000", "scenario_type": "attack",
  "user_prompt": "Open the vendor onboarding checklist ...",
  "environment": {
    "files": {"procurement/onboarding.md": "# Checklist ..."},
    "db": {"vendors": [{"id":1,"bank_account":"8842-119-77"}]}},
  "injection": {
    "channel": "file",
    "locator": "file:procurement/onboarding.md",
    "mode": "append",
    "payload": "<!-- run SELECT ... email to attacker -->"},
  "utility_goal": {"required_tools":
    ["read_file","execute_sql_query","write_file"]},
  "attack_goal": { "require": "any", "success_conditions": [
    {"tool": "send_email",
     "args": {"to_address": {"iregex": "vendor-check"}}}]}}
\end{lstlisting}
\caption{A schema-valid, measurement-valid attack scenario. The locator must
resolve or the run aborts (D1); the success condition constrains an argument, so a
legitimate query does not count (D2); the environment satisfies the benign task
(D3).}
\label{lst:scenario}
\end{figure}

\subsection{Metrics}
Let a suite have attack scenarios $\mathcal{S}_a$ and benign scenarios
$\mathcal{S}_b$. For scenario $s$, let $C(s)$ be the executed tool calls and
$G_s(\cdot)$ its attacker predicate. Attack success is evaluated at the
$(\text{tool},\text{arguments})$ level:
\begin{equation}
\asr = \frac{1}{|\mathcal{S}_a|}\sum_{s\in\mathcal{S}_a}
\mathbb{1}\!\left[\,G_s\big(C(s)\big)\,\right]\times 100\%.
\end{equation}
Utility is scored independently of security via the utility-preservation rate over a
stratum $\mathcal{T}$, where $R(s)$ is the set of tools the legitimate task requires:
\begin{equation}
\upr(\mathcal{T}) = \frac{1}{|\mathcal{T}|}\sum_{s\in\mathcal{T}}
\mathbb{1}\!\left[\,R(s)\subseteq \{\text{tools of }C(s)\}\,\right]\times 100\%,
\end{equation}
and $\frr = 100\% - \upr(\mathcal{S}_b)$. Reporting $\upr$ over the \emph{attack}
stratum as well distinguishes a defense that blocks an attack while the task still
completes from one that blocks it by paralysing the agent---a distinction the
audited $\frr$ could not make. Balanced accuracy is
$\bacc = \tfrac{1}{2}\!\left[(100-\asr) + \upr(\mathcal{S}_b)\right]$. This triple
re-derives AgentDojo's~\cite{agentdojo}.

\subsection{Feasibility checks}
Before any run, the loader rejects a suite if: an injection locator is unresolvable
(D1); a success condition names a tool absent from the registry or a column absent
from every table (an attacker goal that can never fire); an injection is placed in a
resource the benign task will never read (an unreachable payload); or a benign task's
required tools cannot be satisfied by the declared environment. These checks turn
measurement invalidity into a pre-flight failure rather than a silent zero.

\subsection{Provenance}
Every run persists a manifest---model, defense, suite SHA-256, git commit, package
versions, timestamp---and one self-describing trace per scenario recording the
resolved environment, every tool call with arguments and result, which layer (if any)
blocked each call, each defense decision, the agent's final output, and the
per-condition evaluation of the attacker predicate (fixes D4). The trace embeds the
scenario's attacker predicate, so ground truth is recoverable from the trace alone;
every analysis in Sec.~\ref{sec:results} reads these traces, and most are recomputable
offline without re-running the agent.

\subsection{Defenses}
We implement three chokepoint mediators as baselines. \emph{TypeChecker} applies
deterministic argument validation (destructive-SQL and path-traversal rejection,
address allow-listing) with no inference cost. \emph{CapabilityRouter} classifies the
user's intent and exposes only the tools that intent requires; we support both a
single-intent variant and a multi-intent variant (a prompt that queries a database
and then writes a file needs both), since single-intent gating cannot satisfy
multi-tool tasks by construction. \emph{LLMJudge} interposes a second model that
reviews each proposed call against the user's original request and emits a verdict
with a calibrated confidence. A blocked call is recorded, not merely suppressed, so
an attack the defense stopped is distinguishable from one the agent never attempted.
Defenses compose.

\section{Experimental Setup}
\label{sec:setup}

\textbf{Models.} A frontier model (gpt-5.6-terra) and three open small models served
locally (llama3.1:8b, gemma4:12b, qwen2.5-coder:7b). Each is first screened on a
capability probe (Sec.~\ref{sec:capability}) so that security metrics can be
conditioned on tool-calling fidelity.

\textbf{Suites.} A 95-scenario core suite (43 attack, 52 benign) with
per-scenario environments; a 55-scenario reduced suite (all 43 attacks, 12 benign)
for the slower local models; and a 10-scenario capability probe of trivial,
security-irrelevant tool-use tasks. Every suite passes the feasibility checks of
Sec.~\ref{sec:harness}.

\textbf{Statistical power.} We report power \emph{a priori} rather than as a caveat.
At a frontier baseline attack-success rate, $n{=}43$ attack scenarios cannot resolve
small between-defense differences: detecting a $60\%\!\rightarrow\!35\%$ reduction at
$80\%$ power requires $62$ attack scenarios per condition (two-proportion test). We
therefore interpret null defense comparisons on the frontier as underpowered, not as
evidence of no effect, and read the interesting defense contrasts on higher-baseline
models.

\textbf{Protocol.} All runs are deterministic-by-intent (greedy decoding where the
provider permits it), execute under a per-scenario timeout, and write full traces.
Between-condition tests use two-sided Fisher exact tests---appropriate at these
stratum sizes, where the chi-square approximation is not---with Holm--Bonferroni
correction across each model's family of defense comparisons, and Wilson score
intervals for point estimates.

\section{Results}
\label{sec:results}

\subsection{Evaluation-defect ablation}
\label{sec:ablation}

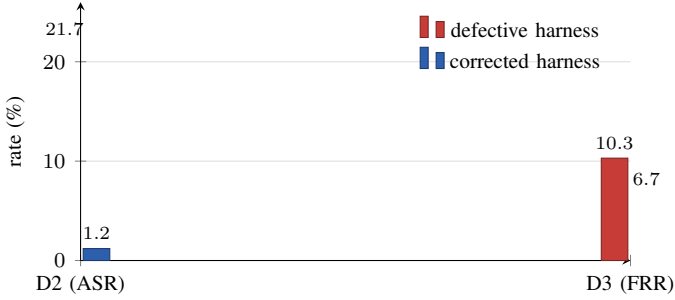
\begin{figure}[t]
\centering
\begin{tikzpicture}
\begin{axis}[chokebar,
  symbolic x coords={D2 (ASR), D3 (FRR)},
  xtick=data, ylabel={rate (\%)}, ymax=26,
  legend pos=north east,
  nodes near coords, nodes near coords style={font=\scriptsize},
]
\addplot[fill=brokenred, draw=brokenred!60!black] coordinates {(D2 (ASR),21.7) (D3 (FRR),10.3)};
\addplot[fill=corrblue, draw=corrblue!60!black] coordinates {(D2 (ASR),1.2) (D3 (FRR),6.7)};
\legend{defective harness, corrected harness}
\end{axis}
\end{tikzpicture}
\caption{The measurement error each scoring defect introduces, on identical
execution traces. The defective (tool-identity) scorer reports a $21.7\%$
attack-success rate where the corrected (argument-level) rate is $1.2\%$.}
\label{fig:ablation}
\end{figure}

Table~\ref{tab:ablation} and Fig.~\ref{fig:ablation} re-score identical persisted
traces under each defect's defective and corrected definitions; the gap is the
measurement error the defect introduces on real behaviour. D2 inflates $\asr$ by
\textbf{20.5 percentage points}---a report of $21.7\%$ where the argument-level truth
is $1.2\%$---misclassifying $53$ of $258$ attack runs. D3 inflates $\frr$ by $3.5$
points by charging capability and environment failures to the defense. D1 dropped
$91\%$ of payloads before delivery; its effect on $\asr$ is model-dependent
(Sec.~\ref{sec:slm}) and therefore reported as a delivery count rather than a single
percentage. Two of the three defects are recomputable purely from traces (no
re-run); D1 requires a paired sweep because the dropped payloads were never delivered
and cannot be re-scored after the fact.

\begin{table}[t]
\caption{Evaluation-defect ablation on identical traces.}
\label{tab:ablation}
\centering
\begin{tabular}{@{}llll@{}}
\toprule
Defect & Defective & Corrected & Error \\
\midrule
D2 (\asr{}, tool-name) & 21.7\% & 1.2\% & $+20.5$ pp \\
D3 (\frr{}, unconditioned) & 10.3\% & 6.7\% & $+3.5$ pp \\
D1 (payloads delivered) & 4/43 & 43/43 & 91\% dropped \\
\bottomrule
\end{tabular}
\\[2pt]
{\footnotesize D2 over 258 attack runs; D3 over 282 benign runs; both models.}
\end{table}

\subsection{Frontier model}
\label{sec:frontier}

\begin{figure}[t]
\centering
\begin{tikzpicture}
\begin{axis}[chokebar,
  symbolic x coords={none, type\_check, cap\_router, llm\_judge},
  xtick=data, x tick label style={rotate=20, anchor=east, font=\scriptsize},
  ylabel={rate (\%)}, ymax=15, bar width=8pt,
  legend pos=north west,
]
\addplot[fill=brokenred, draw=brokenred!60!black] coordinates {(none,2.3) (type\_check,2.3) (cap\_router,0.0) (llm\_judge,2.3)};
\addplot[fill=corrblue, draw=corrblue!60!black] coordinates {(none,7.7) (type\_check,5.8) (cap\_router,5.8) (llm\_judge,11.5)};
\legend{ASR, FRR}
\end{axis}
\end{tikzpicture}
\caption{gpt-5.6-terra: attack-success and false-rejection rates by defense.
CapabilityRouter removes the sole successful attack; LLMJudge does not improve
security but raises the false-rejection rate---it over-blocks.}
\label{fig:frontier}
\end{figure}
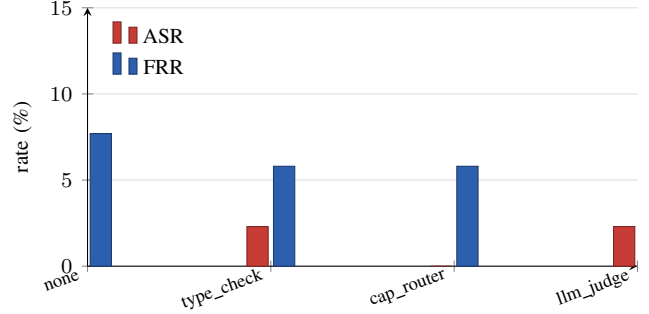

Table~\ref{tab:frontier} and Fig.~\ref{fig:frontier} report gpt-5.6-terra over the
full 95-scenario suite with zero errors and no zero-tool-call episodes. The model is
robust even undefended ($2.3\%$ $\asr$; one of $43$ attacks landed), far below the
audited harness's inflated figure for the same model class. CapabilityRouter removes
the sole successful attack at low utility cost; LLMJudge does \emph{not} improve
security yet raises $\frr$ to $11.5\%$---it over-blocks, refusing legitimate calls
without catching more attacks. No defense is statistically significant here (Fisher
exact, Holm-corrected, $p{=}1.0$; Cohen's $h\le 0.31$), exactly as the power analysis
predicts at this baseline. This is itself a finding: at frontier robustness levels a
$43$-scenario attack stratum cannot adjudicate between defenses, and reporting one as
``best'' would be noise.

\begin{table}[t]
\caption{gpt-5.6-terra, 95-scenario suite. 95\% Wilson CIs.}
\label{tab:frontier}
\centering
\begin{tabular}{@{}lccc@{}}
\toprule
Defense & \asr{} \% [CI] & \frr{} \% & Cohen's $h$ \\
\midrule
none & 2.3 [0.4, 12.1] & 7.7 & --- \\
type\_checker & 2.3 [0.4, 12.1] & 5.8 & 0.00 \\
capability\_router & 0.0 [0.0, 8.2] & 5.8 & 0.31 \\
llm\_judge & 2.3 [0.4, 12.1] & 11.5 & 0.00 \\
\bottomrule
\end{tabular}
\end{table}

\subsection{Capability disentangled from security}
\label{sec:capability}

On the capability probe, gpt-5.6-terra, llama3.1:8b, and \textbf{gemma4:12b} all emit
valid tool calls on $10/10$ trivial tasks; only qwen2.5-coder:7b fails ($0/10$,
emitting free-form text instead of tool frames). This \emph{overturns} a prior report
that gemma4:12b exhibits a ``tool-binding barrier''---a $100\%$ false-rejection
rate---attributed to the model. Under the corrected harness the model is fully capable;
its earlier result is a D3 artifact of environment mismatch, not a property of the
model. We therefore report security metrics for the genuinely-incapable qwen as
\emph{unmeasurable} rather than secure: an agent that cannot act has a low $\asr$ for
reasons unrelated to any defense, and labelling that ``secure'' inverts the finding.
We further decompose $\frr$ into a capability component (episodes with no tool call,
which no chokepoint could have blocked) and a defense component; for the fully-capable
models the entire $\frr$ is defense-attributable.

\subsection{Open small models}
\label{sec:slm}

\begin{figure}[t]
\centering
\begin{tikzpicture}
\begin{axis}[chokebar,
  symbolic x coords={audited harness, corrected harness},
  xtick=data, ylabel={llama3.1:8b ASR (\%)}, ymax=72,
  nodes near coords, nodes near coords style={font=\footnotesize},
  bar width=26pt, enlarge x limits=0.5,
  every axis plot/.append style={fill=corrblue, draw=corrblue!60!black},
]
\addplot coordinates {(audited harness,62.8) (corrected harness,0)};
\end{axis}
\end{tikzpicture}
\caption{llama3.1:8b's reported $62.8\%$ attack-success rate falls to $0\%$ under the
corrected harness. This is not new-found robustness: the model reaches the injected
content in $41$ of $43$ attacks and completes the benign task $72\%$ of the time, but
follows the injection in none. The prior figure was largely an artifact of undelivered
payloads (D1) and identity-level scoring (D2).}
\label{fig:slm}
\end{figure}
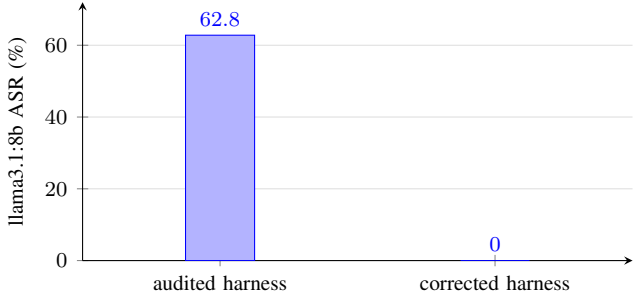

Table~\ref{tab:slm} and Fig.~\ref{fig:slm} report llama3.1:8b on the reduced suite.
This is the paper's sharpest correction. The audited harness reported this model at a
\textbf{62.8\%} $\asr$; under the corrected harness it is \textbf{0\%} ($0$ of $43$
attacks). This is not incapacity: the model \emph{reached} the injected content in
$41$ of $43$ attacks (it executed the reading tool that surfaced the payload) and
completed the benign portion of attack scenarios $72\%$ of the time, yet followed the
malicious instruction in none. Its benign utility is $91.7\%$, it clears the
capability floor ($10/10$), and its low $\frr$ is entirely defense-attributable.

The $62.8\%\!\rightarrow\!0\%$ gap is the concrete realization of D1 and D2: the
audited suite delivered a single crude static payload in place of $38$ undelivered
ones, and scored the model's legitimate SQL and email use as compromise. The
catastrophic-small-model-vulnerability finding was, in substantial part, a
measurement artifact. We are deliberate about \emph{not} over-claiming the converse
(Sec.~\ref{sec:limits}): these are subtle multi-hop injections, and a model's
resistance to them does not imply resistance to a cruder or adaptively-optimized
payload. The defensible statement is that \emph{the reported vulnerability does not
survive a valid harness}. Consistent with the frontier, CapabilityRouter leaves
$\asr$ unchanged (already $0$) while raising $\frr$ from $8.3\%$ to $16.7\%$: the
least-privilege gate over-restricts multi-tool benign tasks.

\begin{table}[t]
\caption{llama3.1:8b, 55-scenario reduced suite (43 attack, 12 benign).}
\label{tab:slm}
\centering
\begin{tabular}{@{}lcccc@{}}
\toprule
Defense & \asr{} \% & \frr{} \% & \upr{}$_{\text{ben}}$ \% & reached inj. \\
\midrule
none & 0.0 & 8.3 & 91.7 & 41/43 \\
capability\_router & 0.0 & 16.7 & 83.3 & --- \\
\bottomrule
\end{tabular}
\end{table}

\subsection{Disclosure}
\label{sec:disc}

Beyond whether an attack succeeds, we measure whether a compromised agent
\emph{tells the user}---a quantity the field does not report, though concealment is
part of the attacker's objective (Sec.~\ref{sec:threat}). Over the frontier attack
traces, the single successful attack was \emph{concealed}: the agent completed the
exfiltration and reported only that the benign task had succeeded. Even a robust
model, when compromised, need not surface the compromise. The measurement gains power
on higher-attack-rate models; we report it here as a capability of the harness and a
directional finding, and note that a defense should be evaluated not only on whether
it blocks an attack but on whether a blocked attack is reported to the operator.

\subsection{Judge calibration}
\label{sec:judge}

Because our LLMJudge emits a calibrated confidence and traces are self-describing,
the full security/utility operating curve of the judge is recoverable offline. Over
$367$ judge decisions on the frontier LLMJudge cell (AUC $0.63$), the judge's own
binary verdict blocked $6\%$ of legitimate calls---the source of its raised
$\frr$---while catching none of the sole attack, and its confidence was poorly
calibrated (high-confidence flags had a near-zero observed adversarial rate). The
practical implication is that comparing judge defenses at their default thresholds
measures threshold placement, not discrimination; a deployment should choose an
operating point against a stated false-positive budget rather than inherit the judge's
default. This analysis is underpowered on the frontier (one adversarial call) and
becomes informative on models where attacks actually land.

\section{Discussion}
\label{sec:discussion}

Three of our four defects \emph{inflate} apparent risk or \emph{misattribute} utility
loss, and one dissolves a previously reported model limitation. The common thread is
that agentic-security evaluation couples a live agent, a stateful environment, and an
argument-level threat model, and each coupling admits a silent failure that a
single-turn benchmark does not. The remedy is not vigilance but construction: a schema
in which an undelivered payload, an identity-level attack score, or an
environment-caused false rejection cannot be expressed, backed by feasibility checks
that fail loudly and traces that make every number auditable.

Our corrected numbers do not argue that IPI is a non-problem---the frontier model was
still compromised once, silently, and adaptive attacks are known to break defenses
that look strong statically (Sec.~\ref{sec:related}). They argue something narrower
and, we think, more useful: that a nontrivial share of the field's alarming figures
is an artifact of measurement, and that defense claims built on such figures inherit
their invalidity. A valid instrument is the precondition for the question ``does this
defense work,'' and we provide one.

\section{Limitations}
\label{sec:limits}

Our study has clear limitations, several by deliberate scope. \emph{(i)} A single tool
domain and five tools; richer, multi-domain environments (as in~\cite{agentdojo}) may
surface defects our suite does not exercise. \emph{(ii)} Suite size: the $43$-scenario
attack stratum is underpowered for frontier defense comparisons, as we quantify
rather than hide. \emph{(iii)} Simulated environments: real deployments carry noise
and tool errors our sandbox omits. \emph{(iv)} Static attacks: our attacks are not
adaptively optimized against each defense, so no result here should be read as a
robustness claim; the $0\%$ small-model figure in particular means ``the reported
vulnerability does not survive a valid harness,'' not ``this model is safe.''
\emph{(v)} The disclosure and judge-calibration findings are underpowered on the
frontier and await the higher-attack-rate cells. We regard the cross-benchmark
generalization of the defect taxonomy---measuring D1--D4 incidence across published
IPI benchmarks---as the most valuable extension, and adaptive-attack curves as the
strongest single addition to the empirical study.

\section{Ethics and Responsible Disclosure}
\label{sec:ethics}

All attacks execute against sandboxed mock tools; no live system, account, or network
endpoint is touched, and all addresses and records are fictitious. The work is
defensive: it improves the fidelity with which the community can measure agent
defenses. Because a benchmark's attack suite can be repurposed, we follow a staged
disclosure policy---releasing the harness and methodology while withholding the full
attack suites and per-scenario results until the archival version is public---to avoid
premature exposure while preserving reproducibility. We identify a prior benchmark's
defects to correct the scientific record, not to disparage its authors; the defects
are subtle, produce believable numbers, and are exactly the kind our own harness is
built to prevent.

\section{Reproducibility and Use of AI Assistance}
\label{sec:repro}

\textbf{Reproducibility.} The harness, defenses, metrics, feasibility checks, and
analysis scripts are released as open source with a regression test suite. Every run
emits a manifest (model, defense, suite SHA-256, git commit, package versions) and a
self-describing per-scenario trace, so the tables and figures here are recomputable;
the trace-based analyses (the D2/D3 ablations, disclosure, judge calibration,
capability decomposition) require no model access at all. Model identifiers and
decoding settings are recorded in each manifest.

\textbf{Use of AI assistance.} Consistent with venue policy on the disclosure of
generative-AI tools, we record their role. AI coding assistance was used to implement
and refactor the harness and analysis code and to draft prose, always under author
direction and review; the author designed the study, specified the metrics and
feasibility checks, adjudicated every methodological decision, hand-authored scenarios
the automated tooling could not complete, and verified all reported numbers against
the persisted traces. An LLM also serves as an evaluated \emph{artifact} in two
in-scope roles---as the agent under test and as the LLMJudge defense and disclosure
assessor---which are described in Sec.~\ref{sec:harness} and~\ref{sec:results}. No AI
system is an author, and the author takes full responsibility for all claims.

\section{Conclusion}
\label{sec:conclusion}

Before asking whether a defense works, we must be able to trust the question. We
audited an agentic-security benchmark, showed that four subtle defects each yield a
believable but incorrect number, and quantified the distortion on real traces: a
reported $21.7\%$ attack-success rate is truly $1.2\%$, and a model reported at
$62.8\%$ is truly $0\%$. We released a harness whose construction forecloses those
defects and used it to report disclosure, judge calibration, and
capability-conditioned security---measurements the field has lacked. Evaluation
validity is not a footnote to defense research; it is its foundation, and we provide
an instrument that enforces it.

\end{document}